\documentclass[10pt,aps,pra,twocolumn]{revtex4-2}
\usepackage{amsmath}
\usepackage{amssymb}
\usepackage{graphicx}
\usepackage{hyperref}
\usepackage[english]{babel}
\usepackage{babel}
\usepackage[usenames,dvipsnames]{color}

\newcommand{\bear}{\begin{eqnarray}}
\newcommand{\eear}{\end{eqnarray}}
\newcommand{\be}{\begin{equation}}
\newcommand{\ee}{\end{equation}}
\newcommand{\beqn}{\begin{eqnarray}}
\newcommand{\eeqn}{\end{eqnarray}}
\newcommand{\beqnn}{\begin{eqnarray*}}
\newcommand{\eeqnn}{\end{eqnarray*}}
\begin{document}

\title{Dynamical Casimir effect via modulated Kerr or higher nonlinearities}
\author{A. V. Dodonov and V. V. Dodonov}
\email{adodonov@unb.br, vdodonov@fis.unb.br}
\affiliation{Instituto de F\'{\i}sica, Universidade de Bras\'{\i}lia, Caixa Postal 04455, CEP
70919-970 Bras\'{\i}lia, DF, Brasil
\\
International Centre of Physics, University of Brasilia, 70297-400 Brasilia,  Federal District, Brazil
}

\begin{abstract}
We show two examples in which the dynamical Casimir effect can be achieved
by modulating the Kerr or higher order nonlinearities. In the first case the
cavity field is coupled to an arbitrary number of qubits or an harmonic
oscillator via the dipole interaction. In the second case, the modulation of
the nonlinearities is accompanied by the off-resonance modulation of the cavity frequency.
 We present the analytic description of the
phenomenon and supplement it with numeric simulations, demonstrating that photons can be created from vacuum and
the resulting  hyper-Poissonian photon statistics is very different from the squeezed vacuum
state.
\end{abstract}

\maketitle

\section{Introduction}

A possibility of creating quanta of the electromagnetic field from the
initial vacuum state in cavities with moving boundaries, first predicted by
Moore \cite{moore} and called nowadays as the dynamical Casimir effect
(DCE), was a subject of numerous studies for several decades (see, e.g., the
reviews \cite{rev1,rev2,rev3,rev4,rev5}). However, till now it seems
impossible to observe the effect in its \textquotedblleft
pure\textquotedblright\ form, due to small velocities (compared with the
speed of light) of real boundaries that could be achieved in a laboratory.
Therefore, the idea of \emph{simulating\/} this motion in more simple
arrangements, resulting in the parametric amplification of vacuum
fluctuations was considered for a long time by many authors \cite%
{Yabl89,Yabl89-2,Man91,Oku95,Loz,Padua,Zhao13mag,Hizh16}. Such phenomena can
be called as the \emph{parametric DCE\/} (or PDCE). One of possibilities is
to use some electrical circuits (waveguides) with distributed or lump
elements, whose parameters (e.g., capacitance, inductance, magnetic flux,
critical current, etc.) could be made time-dependent \cite%
{Man91,Seg07,Fujii11,Berdi14}. The idea to use a superconducting coplanar
waveguide in combination with a Josephson junction was developed in \cite%
{AVD09,Johan09,Johan10,Wilson10}, and the experiments were reported in \cite%
{Wilson11-Nature,Johan13,Paraoanu13,Svensson18}. Further improvements of
experimental schemes were suggested in \cite%
{Wust13,Rego14,Doukas15,Corona16,Lombardo16,Sabin17superlum,Lombardo18,Bosco19,Ma19}%
. In particular, the circuit QED with \textquotedblleft artificial
atoms\textquotedblright\ (qubits) was the subject of studies \cite%
{AVD14,AVD15,Felicetti15,AVD15souza,Silva16,Rossatto16,AVD16,Gu17,Zhukov17,Zhukov18,AVD18,Wust19}%
.

Although the simplest models predict an exponential growth of the number of
quanta created from vacuum under the DCE parametric resonance conditions
\cite{rev1}, realistic numbers can be limited due to many factors. One of
such factors is related to unavoidable nonlinearities in real systems \cite%
{Srivast06,Roman17,Paredes19,Akop21,Trun21}.
As a rule, nonlinearities with time-independent parameters play a negative role,
leading the system out the resonance, thus
diminishing the number of quanta that could be created from vacuum.
The aim of our paper is to show that {\em time-modulated\/} nonlinear effects  can
be used to create photons
from vacuum, resulting in new quantum states of the electromagnetic field (quite
different from the typical squeezed vacuum state
generated in the linear parametric amplification processes).
Namely, we assume that
the Hamiltonian has the form (hereafter $\hbar =1$ and $\eta >0$)
\begin{equation}
\hat{H}=\varepsilon \sin (\eta t)\hat{n}^{k}+\hat{H}_{0},  \label{ham}
\end{equation}%
where $\hat{n}=\hat{a}^{\dagger }\hat{a}$ is the photon number operator ($%
\hat{a}$ and $\hat{a}^{\dagger }$ being the standard annihilation and
creation operators) and $k\geq 2$ is an integer (the case $k=1$,
corresponding to the modulation of the cavity frequency, was thoroughly
studied previously in \cite{AVD15souza,Silva16}). $\varepsilon$ is the amplitude of modulation of the nonlinearity, while $\eta$ is the modulation frequency. We consider two examples
of the \textquotedblleft bare\textquotedblright\ Hamiltonian $\hat{H}_{0}$.
The first one (Sec. \ref{sec-qubits}) describes the interaction of the field
mode with a chain of qubits or a harmonic oscillator, whereas the second
one corresponds to a non-resonantly modulated cavity (Sec. \ref{sec-modcav}%
). A discussion of results is made in Sec. \ref{sec-concl}.

\section{Employing dispersive qubits}

\label{sec-qubits}

In this section, we consider the bare Hamiltonian $\hat{H}_{0}$ describing
the \emph{quantum Dicke model} \cite{dicke,HP} including a Kerr nonlinearity
with a constant strength $\alpha $ (whose role is to minimize the
qubit-induced nonlinearity, as will be seen shortly), $\hat{H}_{0}= \omega
\hat{n} + \alpha \hat{n}^{2} + \hat{H}_1$, where
\begin{equation}
\hat{H}_{1}= \sum_{l=1}^{N}\left[ \frac{\Omega }{2}\hat{\sigma}_{z}^{(l)}+g(%
\hat{a}+\hat{a}^{\dagger })(\hat{\sigma}_{+}^{(l)}+\hat{\sigma}_{-}^{(l)})%
\right].  \label{H0}
\end{equation}%
Here $\omega$ is the constant cavity frequency, $\Omega$ is the
constant atomic transition frequency, $g$ is the atom-field coupling constant
and $N$ is the number of identical noninteracting atoms. The qubit operators are
$\hat{\sigma}_{-}^{(l)}=|g^{(l)}\rangle \langle e^{(l)}|$, $\hat{\sigma}%
_{+}^{(l)}=|e^{(l)}\rangle \langle g^{(l)}|$ and $\hat{\sigma}%
_{z}^{(l)}=|e^{(l)}\rangle \langle e^{(l)}|-|g^{(l)}\rangle \langle g^{(l)}|$%
, where $|g^{(l)}\rangle $ and $|e^{(l)}\rangle $ denote the\ ground and
excited states of the $l$-th qubit, respectively.
The Hamiltonian (\ref{H0}) is the starting point in the studies
devoted to the interaction of a single mode of the electromagnetic field with
ensembles of two-level objects (``qubits'').

To obtain a closed analytical description we employ the normalized \emph{%
Dicke} states with $k$ atomic excitations (denoted by bold index) \cite{dicke,HP}
\begin{equation*}
|\mathbf{k}\rangle =\sqrt{\frac{k!(N-k)!}{N!}}\sum_{p}|e^{\left( 1\right)
}\rangle \cdots |e^{\left( k\right) }\rangle |g^{\left( k+1\right) }\rangle
\cdots |g^{N}\rangle ,
\end{equation*}%
where the sum runs over all allowed permutations of excited and non-excited
qubits and $k=0,1,\ldots ,N$. In terms of the collective qubit operators $%
\hat{\sigma}_{k,j}\equiv |\mathbf{k}\rangle \langle \mathbf{j}|$ we have
\begin{equation}
\hat{H}_{1}=\sum_{k=0}^{N}\left[ k\Omega \hat{\sigma}_{k,k}+gf_{k}(\hat{a}+%
\hat{a}^{\dagger })(\hat{\sigma}_{k,k+1}+\hat{\sigma}_{k+1,k})\right] ,
\label{H1N}
\end{equation}%
where $f_{k}\equiv \sqrt{(k+1)(N-k)}$. Hamiltonian (\ref{H1N}) can be simplified if $N\gg 1$ and the maximum number
of atomic excitations is small compared to the number of atoms ($k_{\max }\ll N$).
Writing $g =g_{ho}/\sqrt{N}$ and taking the limit $N\to\infty$ one can arrive
at the Hamiltonian of two coupled harmonic oscillators,
\begin{equation}
\hat{H}_{1}= \Omega \hat{b}^{\dagger }\hat{b}+g_{ho}(\hat{a}+\hat{a}%
^{\dagger })(\hat{b}+\hat{b}^{\dagger }),
\label{H1inf}
\end{equation}%
where $\hat{b}=\sum_{k=0}^{\infty }\sqrt{k+1}\hat{\sigma}_{k,k+1}$ is the
collective atomic operator, satisfying the standard bosonic commutation
relation $[\hat{b},\hat{b}^{\dagger }]=1$.

In the presence of the nonlinear term in the total  Hamiltonian $\hat{H}$ (\ref{ham}),
it is convenient
\cite{AVD18,AVD19} to expand the wavefunction  as
\begin{equation}
|\psi \rangle =\sum_{n}\exp \left[ (i\varepsilon \xi_n/\eta)
\cos (\eta t) -it\lambda _{n}\right] c_{n}(t)|\varphi _{n}\rangle,
\label{ttt}
\end{equation}%
where $\xi_n = \langle \varphi _{n}|\hat{n}^{k}|\varphi _{n}\rangle$ and $%
|\varphi _{n}\rangle $ is the eigenstate (dressed-state) of $\hat{H}_{0}$
with the eigenvalue $\lambda _{n}$; the sum runs over all the dressed-states, with the index $n$ increasing with energy (i.e., $\lambda_{n+1}\ge\lambda_{n}$). After substituting Eq. (\ref{ttt}) into the Schr\"odinger equation, one finds that the probability amplitudes of the
dressed-states, $c_n$, obey the set of differential equations
\begin{equation}
i\dot{c}_{m} = 2\sin (\eta t)\sum_{n\neq m}\exp \left[ iQ_{mn}\cos (\eta t)
-it \lambda_{nm} \right] R_{mn}c_{n},
\label{a}
\end{equation}
where $\lambda_{nm} = \lambda _{n}-\lambda _{m}$ and
\begin{equation}
Q_{mn}\equiv (\varepsilon /\eta)\left( \xi_n - \xi_m \right), \;\;
R_{mn}\equiv (\varepsilon /2)\langle \varphi_{m}|\hat{n}^{k}|\varphi_{n}\rangle.
\label{d}
\end{equation}%

Using the Jacobi-Anger expansion
\[
\exp[iz\cos(x)] = J_0(z) + 2\sum_{l=1}^{\infty } i^l J_l(z)\cos(nx)
\]
together with the recurrence relation
\[
J_{l-1}\left( z\right) +J_{l+1}\left( z\right)
=2lz^{-1}J_{l}\left( z\right),
\]
 one can rewrite
  Eq. (\ref{a}) in terms of
the Bessel functions of the first kind: 
\begin{equation}
\dot{c}_{m}=-4\sum_{n\neq m}R_{mn}c_{n}e^{-it \lambda_{nm}
}\sum_{l=1}^{\infty }li^{l}\frac{J_{l}\left(Q_{mn}\right)}{Q_{mn}} \sin (l\eta t).
\label{ty}
\end{equation}
The equivalent sets of equations (\ref{a}) and (\ref{ty}) are exact but rather complicated.
However, they can be simplified significantly under the condition $|Q_{mn}|\ll 1$
(which is satisfied for typical experimental situations).
The RHS of Eq. (\ref{ty}) exhibits
fast oscillations as function of time, which average to zero, unless the
modulation frequency assumes resonant values $\eta \approx |\lambda _{nm}|/l$
simultaneously with nonzero $R_{mn}$. In this case, the external perturbation
drives the transition between the dressed-states $\{\varphi _{m},\varphi
_{n}\}$; since the modulation frequency is controlled externally, the order $%
l$ of the resonance and the coupled states are chosen by the
experimentalist. In the realistic scenario $lJ_{l}\left( Q_{mn}\right)
R_{mn}/Q_{mn}\ll l\eta +\left\vert \lambda _{nm}\right\vert $ for the
relevant values of $m$, $n$ and $l$,  one can neglect rapidly oscillating
terms to obtain%
\[
\dot{c}_{m}=\sum_{n\neq m}\theta _{mn}R_{mn}c_{n}\sum_{l=1}^{\infty }li^{l-1}%
\frac{2J_{l}\left( Q_{mn}\right) }{Q_{mn}}e^{it\theta _{mn}\left( \left\vert
\lambda _{nm}\right\vert -l\eta \right) },  
\]
where $\theta _{mn}=\mbox{sign}(\lambda _{mn})$.  Since $J_{n}\left( x\right)
\approx x^{n}/(2^{n}n!) $ for $|x| \ll 1$, we have $2lJ_{l}\left( Q_{mn}\right)
/Q_{mn}\approx Q_{mn}^{l-1}/[2^{l-1}(l-1)!]$ if $|Q_{mn}|\ll 1$. Therefore, the benefit of using
higher order resonances is offset by much lower transition rates \cite%
{Silva16}.  For simplicity, here we focus on the lowest order resonance $l=1$.
In this case,
the exact set of equations can be replaced with a simpler one,
\begin{equation}
\dot{c}_{m}=\sum_{n\neq m}\theta _{mn}R_{mn}c_{n}\exp\left[it\theta _{mn}\left( \left\vert
\lambda _{nm}\right\vert -\eta \right) \right],
\label{mi}
\end{equation}%
Only the terms
satisfying the resonance condition
$ |\lambda _{nm}| \approx \eta$ simultaneously with nonzero $R_{mn}$ make the main contribution
to the right-hand side of this equation.
We also notice that the above approximation introduces small shifts \cite%
{AVD14} to the resonant modulation frequency $|\lambda _{nm}|$, which are
more easily found numerically.

For the scope of this work it is sufficient to work in the dispersive regime
and weak Kerr nonlinearity: $gf_{k}\sqrt{n}$, $2|\alpha |n\ll |\omega
-\Omega |$ for all relevant values of $n$ and $k$. In this regime one can
find the eigenstates of $\hat{H}_{0}$ via the standard nondegenerate
perturbation theory. Since DCE concerns the generation of photon pairs from
vacuum, we only need the dressed-states in which the atoms remain
approximately in the collective ground state $|\mathbf{0}\rangle $. To the
second order in $g$ these (non-normalized) eigenstates read%
\begin{eqnarray}
|\varphi _{n}\rangle &=&|\mathbf{0},n\rangle +\frac{g\sqrt{Nn}}{\omega
-\Omega }|\mathbf{1},n-1\rangle -\frac{g\sqrt{N(n+1)}}{\omega +\Omega }|%
\mathbf{1},n+1\rangle \notag \\
&&+\frac{Ng^{2}\sqrt{n(n-1)}}{2\omega (\omega -\Omega )}|\mathbf{0}%
,n-2\rangle \notag\\
&&+\frac{g^{2}N\sqrt{(n+1)(n+2)}}{2\omega (\omega +\Omega )}|\mathbf{0}%
,n+2\rangle \label{weight}\\
&&+\frac{g^{2}\sqrt{2Nn(N-1)(n-1)}}{2(\omega -\Omega )^{2}}|\mathbf{2}%
,n-2\rangle \notag\\
&&+\frac{g^{2}\sqrt{2N(N-1)}\left[ \omega -\Omega (2n+1)\right] }{2\Omega
(\omega ^{2}-\Omega ^{2})}|\mathbf{2},n\rangle \notag\\
&&+\frac{g^{2}\sqrt{2N(N-1)(n+1)(n+2)}}{2(\omega +\Omega )^{2}}|\mathbf{2}%
,n+2\rangle. \notag
\end{eqnarray}%
The dressed-states with excited atoms can be found similarly, but they are
not important here.

The energy difference between the dressed states
differing by (roughly) two photons reads%
\begin{gather*}
\frac{\lambda _{n+2}-\lambda _{n}}{2}\approx \omega +2\alpha +2n\left[
\alpha +\frac{2\Omega g^{4}N(\omega ^{2}+3\Omega ^{2})}{(\Omega ^{2}-\omega
^{2})^{3}}\right] \\
-\frac{2\Omega Ng^{2}}{\Omega ^{2} \!-\! \omega ^{2}}\left[ 1-g^{2}\frac{3\omega
(\omega ^{2} \!+\! 3\Omega ^{2})+\Omega \lbrack (5N \!-\! 8)\omega ^{2}-N\Omega ^{2}]}{%
\omega (\Omega ^{2} \!-\! \omega ^{2})^{2}}\right].
\end{gather*}%
By adjusting the static value of the Kerr nonlinearity to
\begin{equation*}
\alpha _{0}=-\frac{2\Omega g^{4}N(\omega ^{2}+3\Omega ^{2})}{(\Omega
^{2}-\omega ^{2})^{3}},
\end{equation*}%
 this energy difference becomes independent of the photon number to the 4th order in $g$:
\begin{eqnarray}
&&\frac{\lambda _{n+2}-\lambda _{n}}{2}\approx \omega -\frac{2\Omega Ng^{2}}{%
\Omega ^{2}-\omega ^{2}}  \label{v1} \\
&&\times \left[ 1-g^{2}\frac{\omega (\omega ^{2}+3\Omega ^{2})+\Omega
\lbrack (5N-8)\omega ^{2}-N\Omega ^{2}]}{\omega (\Omega ^{2}-\omega ^{2})^{2}%
}\right].  \notag
\end{eqnarray}%
Therefore, all the low lying dressed-states (satisfying $gf_{k}\sqrt{n}\ll
|\omega -\Omega |$) {\em can be coupled resonantly by a single-tone modulation}.
This means a possibility of the photon creation from vacuum via the modulation of nonlinearities.
However, the dynamics of this process is different from the standard DCE
(when the time-dependent interaction Hamiltonian is proportional to  $\hat{a}^{\dagger 2}+\hat{a}^{2}$ \cite{law}),
as well as from the case when the photon generation can be achieved via the modulation of parameters $\Omega$ or $g$
\cite{AVD09,AVD14,AVD15,AVD15souza,ibe}.
To show the origin of the difference, let us analyze the structure of matrix elements (\ref{d}) determining the evolution
through the set of equations (\ref{a}).
Under the assumptions made before, these matrix elements can be written as follows,
\begin{equation*}
Q_{n,n+2}\approx (\varepsilon /\eta )\left[ (n+2)^{k}-n^{k}\right],
\end{equation*}%
\be
R_{n,n+2}
=\frac{N\varepsilon (g/\omega )^{2}}{2\left(\nu^2 -1\right)}
\sqrt{(n+1)(n+2)}{M}_{k}(n,\nu),
  \label{vf}
\ee
\be
{M}_k(n,\nu) = (n+1)^k - \frac{1-\nu}{2}(n+2)^k - \frac{1+\nu}{2} n^k,
\label{li}
\ee
where $\nu = \Omega/\omega$.
For the realistic case $2kn_{\max }^{k-1}\varepsilon \ll \eta $ (where $n_{\max }$ is the
maximum relevant photon number) one has $Q_{n,n+2}\ll 1$, so
 the actual transition rate between the states $|\varphi_{n}\rangle$ and $|\varphi _{n+2}\rangle $
is given by the coefficient  $R_{n,n+2}$
 [this means, in particular, that Eq. (\ref{mi}) is indeed an excellent approximation to Eq. (\ref{ty})].
 Moreover, since the above results were derived assuming that the
atoms remain predominantly in the ground states (so the condition $k_{\max}\ll N$ is fulfilled), we can immediately infer
the results for the harmonic oscillator by making the substitution $g=g_{ho}/%
\sqrt{N}$ and then taking the limit $N\rightarrow \infty $ (in this case $%
\alpha _{0}=0$, therefore the static Kerr term is not required).

As shown in previous papers \cite{AVD09,AVD14,AVD15,AVD15souza,law,ibe}, %
  the transition rate $R_{n,n+2}$ scales as $\sqrt{(n+1)(n+2)}$
for the modulation of parameters $\Omega $ or $g$, as well as
in the standard DCE.
On the other hand, Eq. (\ref{vf}) contains an extra factor $M_k(n,\nu)$ (\ref{li}).
Only ${M}_1 = 2\nu$ does not depend on the quantum number $n$.
This situation corresponds to the ``standard DCE case'' (i.e., the modulation of the cavity
eigenfrequency). All coefficients ${M}_{k}$ with $k \ge 2$ depend on $n$.
In particular,
\[
{M}_{2}= 2(n+1)\nu -1, \quad
{M}_{3}=(3n^{2}+6n+4)\nu -3(n+1).
\]
Therefore, one can expect that the dynamics in the case of $k \ge 2$ can be different from the cases studied earlier.
This conjecture is confirmed numerically in the following subsection.

\subsection{Numeric results for a single qubit}

\begin{figure}[hbt]
\begin{center}
\includegraphics[width=0.49\textwidth]{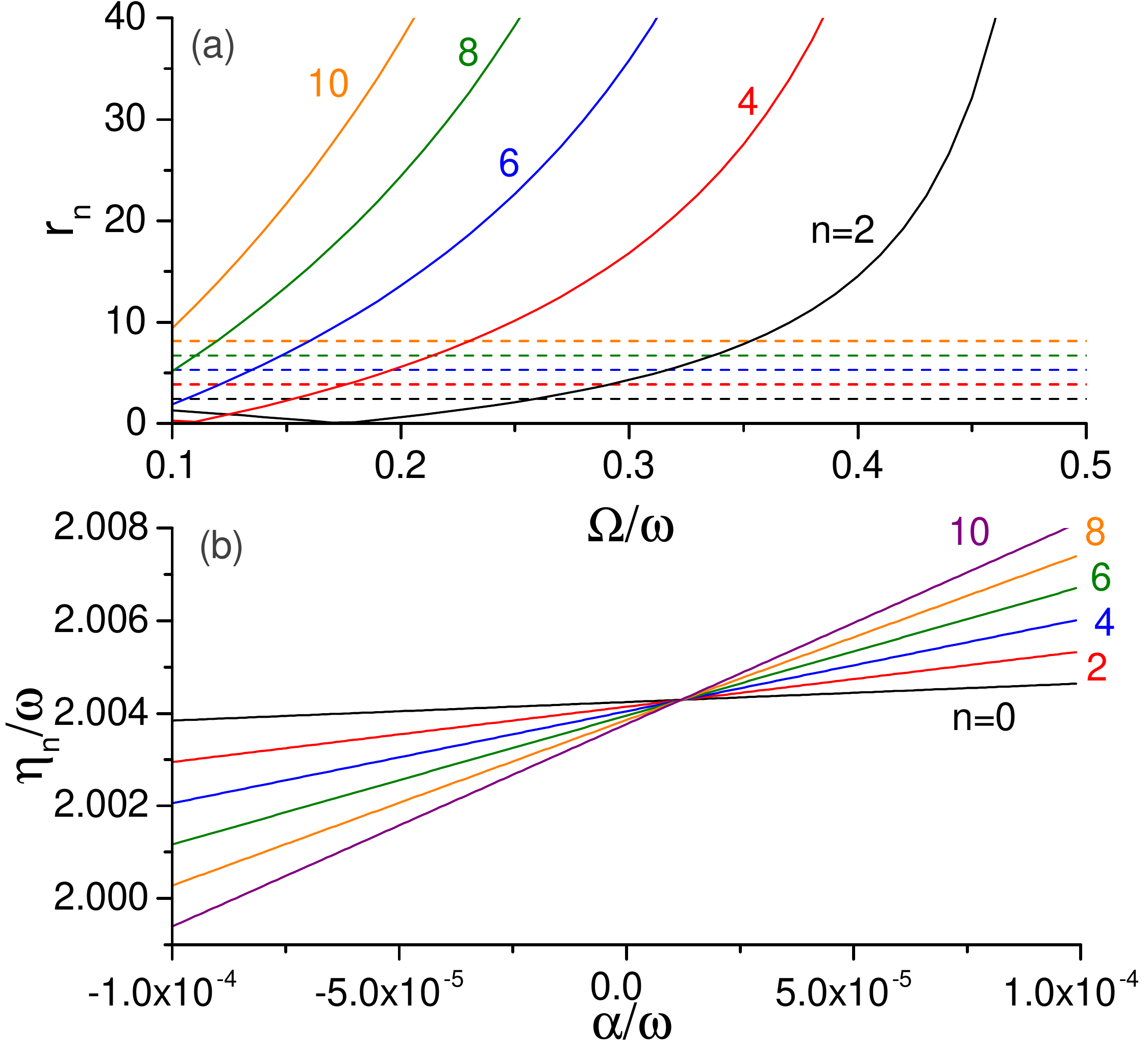} {}
\end{center}
\caption{(Color online) a) Ratio of the transition rates $%
r_{n}=R_{n,n+2}/R_{0,2}$ as function of $\Omega/\omega $. Solid lines indicate $%
r_{n}$ for $k=2$ (modulation of Kerr nonlinearity); dashed lines indicate $%
r_{n}$ for the standard DCE (when $k=1$). b) Energy differences $\protect\eta %
_{n}=\protect\lambda _{n+2}-\protect\lambda _{n}$ between adjacent
dressed-states as functions of the (relative) static Kerr nonlinearity $\alpha/\omega $
for $\Omega /\omega=0.21$. }
\label{fig1}
\end{figure}
The analytic results (\ref{v1}) -- (\ref{li}) were deduced for a weak
atom-field coupling strength $g$, when the transition rate $R_{n,n+2}\propto
(g/\omega )^{2}$ is also small. However, to observe the predicted phenomenon
experimentally, the circuit QED architecture is the most promising candidate,
and there the parameter $g$ can be easily made as large as $0.1\omega $ \cite%
{m1,m2},  while the Kerr
nonlinearity can also be modulated externally in real time \cite{ker}. In
this regime, it is easier to evaluate the transition rate $R_{n,n+2}$ and the
resonant modulation frequencies by diagonalizing numerically the Hamiltonian
$\hat{H}_{0}$, since the system dynamics is still described by Eq. (\ref{ty}).
This is done in Fig. \ref{fig1}.
In Fig. \ref{fig1}a the solid lines
illustrate the ratio of matrix elements $r_{n}\equiv R_{n,n+2}/R_{0,2}$
obtained by exact numeric diagonalization of the Hamiltonian (\ref{H0}) for $%
N=1$ and parameters $k=2$, $\alpha =0$ and $g/\omega =0.07$.
 The condition $gf_{k}\sqrt{n}\ll |\omega -\Omega |$ is not
satisfied in this regime of parameters,
so the approximate expressions (\ref{vf}) and (\ref{li}) do not hold. The
dashed lines illustrate the corresponding ratios for the standard DCE
(i.e., for $k=1$), for
which $r_{n}=\sqrt{(n+1)(n+2)/2}$ does not depend on $\Omega$.
We see that the behavior of $r_{n}$ is
drastically different from the one of standard DCE, therefore, the dynamics
will also be quite different! In Fig. \ref{fig1}b we plot the energy
differences $\eta _{n}=\lambda _{n+2}-\lambda _{n}$ obtained by exact
numeric diagonalization as function of $\alpha /\omega $ for $\Omega /\omega
=0.21$.
For example, typical circuit QED
experiments \cite{Gu17,m2} employ cavity frequencies in the range of 5--15
GHz; the qubit frequencies lie in the range 1--10 GHz and can be tuned by as
much as 1 GHz in 20 ns via external magnetic flux.
We see that for $\alpha /\omega \approx 10^{-5}$ the spectrum
becomes quasi-harmonic, so it should be possible to generate several photons from vacuum
for the modulation frequency $\eta \approx \eta _{0}$.

\begin{figure}[hbt]
\begin{center}
\includegraphics[width=0.49\textwidth]{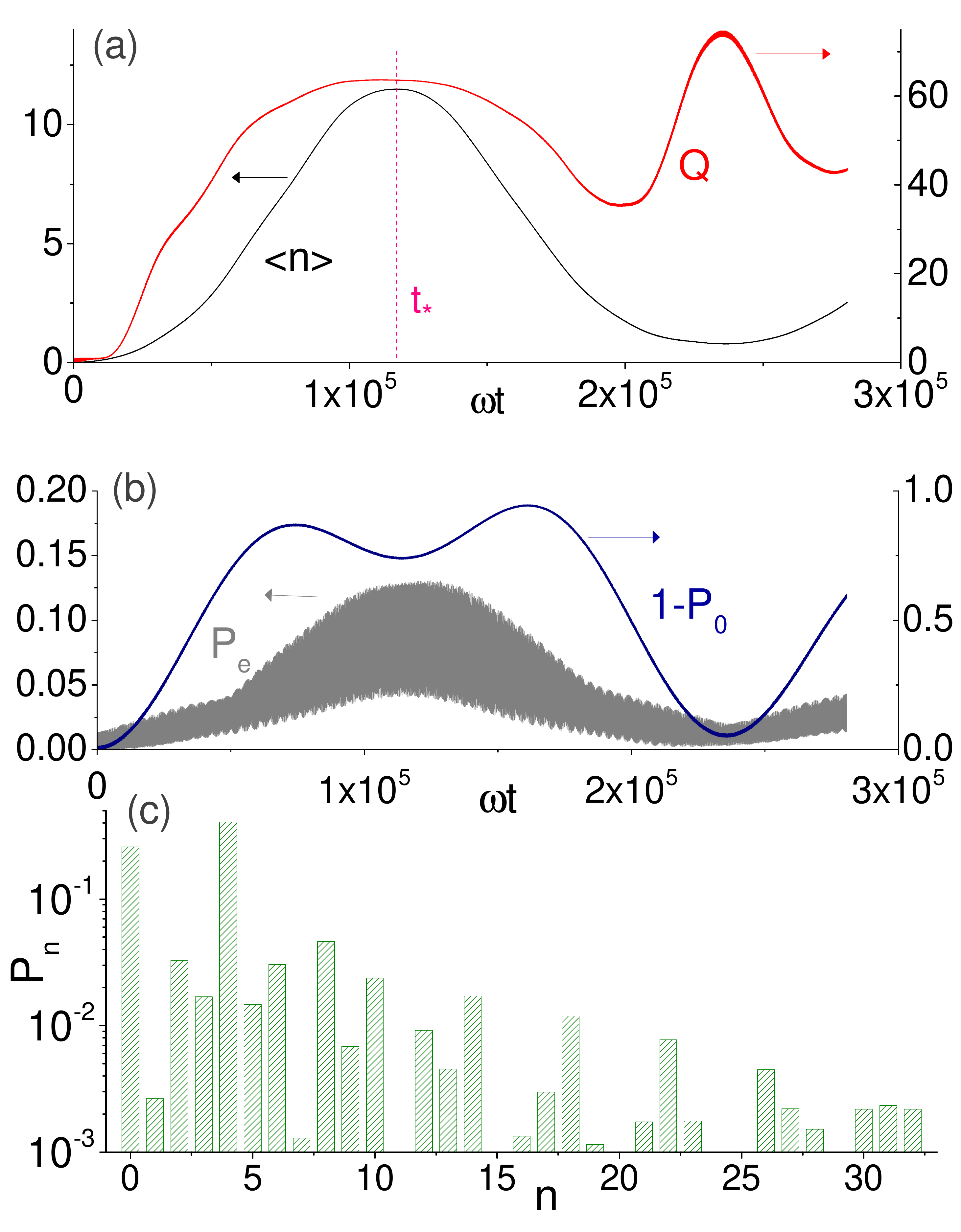} {}
\end{center}
\caption{(Color online) a) Average photon number $\langle n\rangle $ (scale
on left axis) and the Mandel's $Q$-factor (scale of right axis) as function
of time for the modulation of Kerr nonlinearity ($k=2$). b) Dynamics of the
probabilities of atomic excitation $P_{e}$ and nonvacuum photon states $%
1-P_{0}$. c) Photon number statistics at the time instant of maximum $%
\langle n\rangle $ (indicated by $t_{\ast }$ in the panel a).
The values of all fixed parameters are given in the text.
}
\label{fig2}
\end{figure}
Fig. \ref{fig2} shows the dynamics obtained by solving numerically the Schr%
\"{o}dinger equation with the original Hamiltonian (\ref{ham})
 for the initial state $|\mathbf{0},0\rangle
$ and parameters $k=2$, $\Omega /\omega =0.21$, $g/\omega =0.07$, $\alpha
/\omega =10^{-5}$, $\varepsilon /\omega =10^{-2}$ and $\eta /\omega =2.0043$%
. Fig. \ref{fig2}a shows the behavior of the average photon number $\langle
n\rangle $ and the Mandel's factor $Q=[\langle (\Delta n)^{2}\rangle
-\langle n\rangle ^{2}]/\langle n\rangle $ (that quantifies the spread of
the photon number distribution). Fig. \ref{fig2}b shows the behavior of the
atomic excitation probability $P_{e}$ and the probability of occupation of
nonvacuum states of the field: $1-P_{0}$, where $P_{n}=\mathrm{Tr}[|n\rangle
\langle n|\hat{\rho}]$ is the $n$-photon probability and $\hat{\rho}$ is the
total density operator. Fig. \ref{fig2}c shows the photon statistics at the
instant of maximum $\langle n\rangle $ (for the time $\omega t_{\ast
}=1.18\times 10^{5}$). We see that several photons are generated, and the
photon statistics is very different from the squeezed vacuum state that
occurs for the standard cavity DCE. The qubit also becomes slightly excited during the photon generation process; this is easily explained by the fact that the modulation populates $n$-photon dressed-states whose atomic weight is roughly $g^2 n/(\omega-\Omega)^2$
(see Eq. \ref{weight}).

\begin{figure}[htb]
\begin{center}
\includegraphics[width=0.49\textwidth]{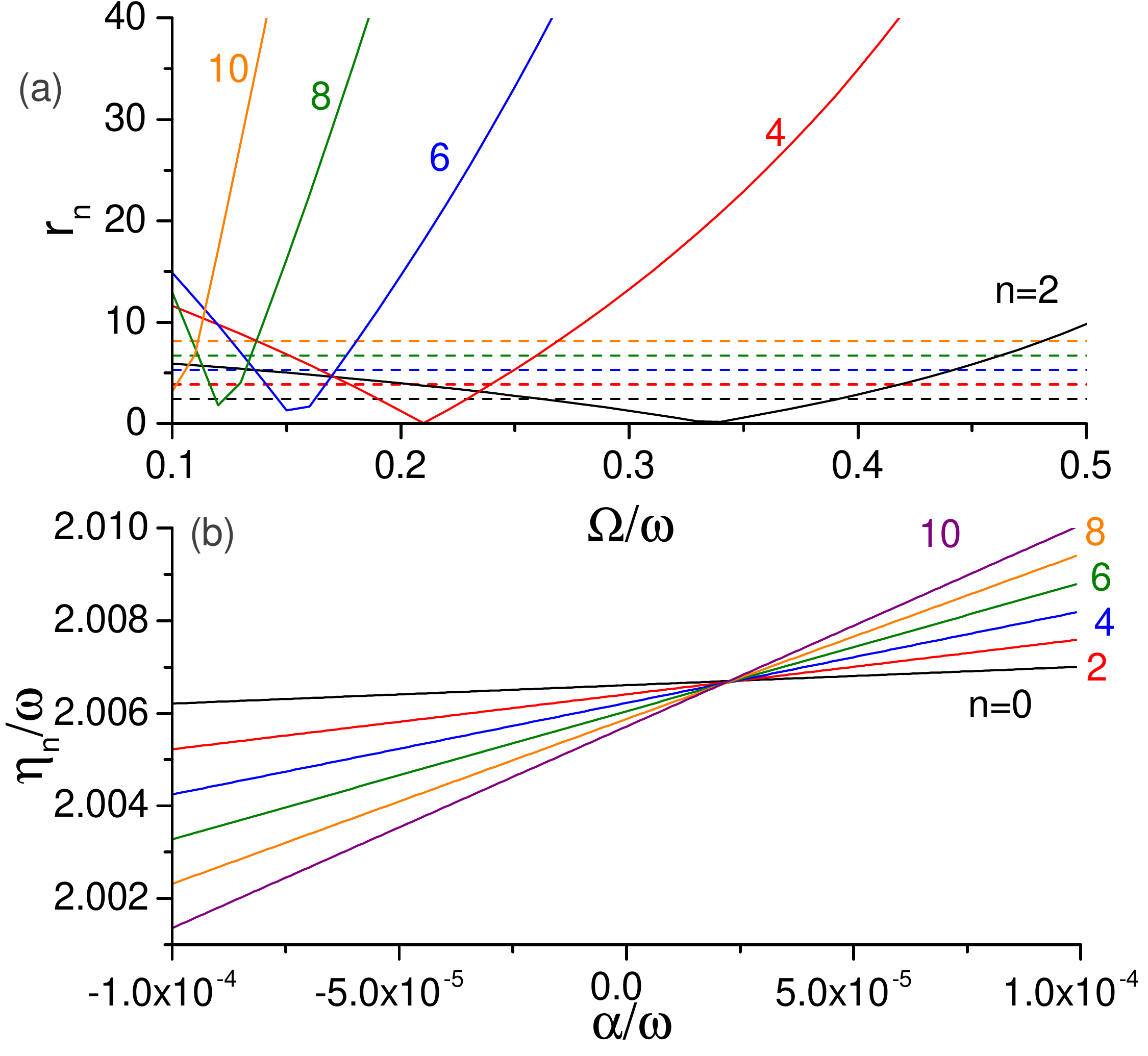} {}
\end{center}
\caption{(Color online) Similar to Fig. \protect\ref{fig1} but for $k=3$
(modulation of the 3-rd order nonlinearity).}
\label{fig3}
\end{figure}
\begin{figure}[htb]
\begin{center}
\includegraphics[width=0.49\textwidth]{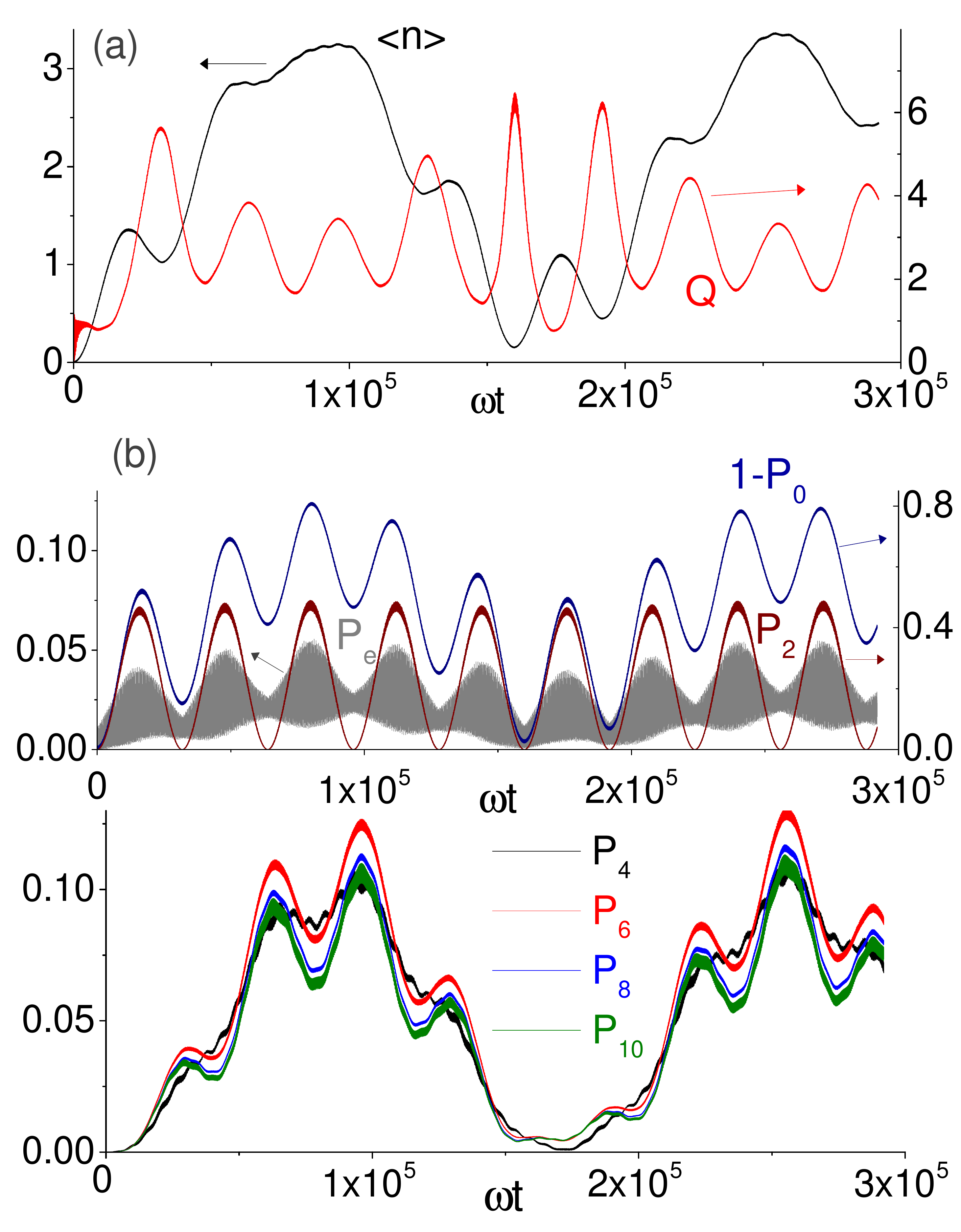} {}
\end{center}
\caption{(Color online) Numeric dynamics for the modulation of the 3-rd
order nonlinearity ($k=3$). a) Average photon number and Mandel's factor. b)
Atomic excitation probability, nonvacuum excitation probability and largest
photon number probabilities as function of time.}
\label{fig4}
\end{figure}
Fig. \ref{fig3} is analogous to Fig. \ref{fig1} but for $k=3$ (all other
parameters are the same). Once again we see that the behavior of $r_{n}$ is
quite different from the typical DCE scenario, and by properly adjusting the static
Kerr nonlinearity $\alpha $ the spectrum can be made quasi-harmonic (in Fig. %
\ref{fig3}b $\Omega /\omega =0.31$). The exact numeric dynamics is shown in
Fig. \ref{fig4} for parameters $k=3$, $\Omega /\omega =0.31$, $g/\omega
=0.07 $, $\alpha /\omega =2.5\times 10^{-5}$, $\varepsilon /\omega =10^{-2}$
and $\eta /\omega =2.0067$. Fig. \ref{fig4}a shows the behavior of $%
\left\langle n\right\rangle $ and $Q$, while Fig. \ref{fig4}b illustrates
the dynamics of $P_{e}$, $1-P_{0}$ and the largest probabilities of
generation of $n$ photons
(other photon-number
probabilities are significantly smaller). In this example, an even number of
up to 10 photons can be generated with significant probabilities.

Figs. \ref{fig2} and \ref{fig4} demonstrate that in the presence of additional subsystems photons
can be generated from vacuum due to the time modulation of cavity
nonlinearities, but the dynamics is completely different from the standard
cavity DCE.

\section{Employing nonresonant cavity modulation}

\label{sec-modcav}

Now we consider a cavity whose frequency is modulated as $\omega (t)=\omega
_{0}+\varepsilon _{\omega }\sin \omega _{1}t$, where $\omega _{1}\neq
2\omega _{0}$, so that the resonant creation of quanta via DCE \cite{rev1}
does not take place. Adding the modulation of Kerr or higher nonlinearities,
we have the Hamiltonian (\ref{ham}) with the standard DCE contribution \cite{law}
\begin{equation}
\hat{H}_{0}=\omega (t)\hat{n}+i\frac{\dot{\omega}}{4\omega }
(\hat{a}^{\dagger 2}-\hat{a}^{2}).
\label{taine}
\end{equation}%
In the interaction picture defined by the unitary transformation
\begin{equation*}
\hat{U}=e^{-iX(t)\hat{n}}~,~X(t)=\frac{\varepsilon _{\omega }}{\omega _{1}}%
(1-\cos \omega _{1}t)+\frac{\omega _{1}t}{2}
\end{equation*}%
the Hamiltonian becomes%
\begin{equation*}
\hat{H}_{0}=\zeta \hat{n}+ \frac{2i\chi\cos \omega _{1}t}{1+(\varepsilon
_{\omega }/\omega _{0})\sin \omega _{1}t}\left(\hat{a}^{\dagger 2}e^{2iX(t)}-\hat{%
a}^{2}e^{-i2X(t)}\right),
\end{equation*}%
where $\zeta =\omega _{0}-\omega _{1}/2$ and $\chi =\varepsilon _{\omega
}\omega _{1}/(8\omega _{0})$. Using the Jacobi-Anger expansion, this
Hamiltonian can be rewritten as a series of the Bessel functions with the argument $%
2\varepsilon _{\omega }/\omega _{1}$.
However, under the realistic condition
$\varepsilon _{\omega }\ll \omega _{0},\omega _{1}$,  we obtain
to the first order in $\varepsilon _{\omega }$
\begin{equation*}
\hat{H}_{0}\approx \zeta \hat{n}+2i\chi \cos \omega _{1}t(\hat{a}^{\dagger
2}e^{i\omega _{1}t}-\hat{a}^{2}e^{-i\omega _{1}t})\,.
\end{equation*}
For low photon numbers $n$, given by the inequality $n\varepsilon _{\omega
}\ll 10\omega _{0}$, one can neglect the rapidly oscillating terms $e^{\pm
2i\omega _{1}t}$ to obtain%
\begin{equation}
\hat{H}_{0}\approx \zeta \hat{n}+i\chi (\hat{a}^{\dagger 2}-\hat{a}^{2}),
\label{Hintfin}
\end{equation}%
Expanding the wavefunction in terms of the eigenstates $|\varphi _{n}\rangle
$ of Hamiltonian (\ref{Hintfin}) as in Eq. (\ref{ttt}), for $|R_{mn}|\ll
\eta $ we obtain%
\begin{equation*}
\dot{c}_{m}\approx \sum_{n\neq m}e^{it\theta _{mn}\left( \left\vert \lambda
_{nm}\right\vert -\eta \right) }\theta _{mn}R_{mn}c_{n}
\end{equation*}%
with the transition rate $R_{mn}$ given by Eq. (\ref{d}).

For the realistic scenario $\chi n\ll |\zeta |$ for all relevant photon
numbers $n$, the (non-normalized) eigenstates can be found from the
nondegenerate perturbation theory as%
\begin{eqnarray*}
|\varphi _{n}\rangle &\approx &|n\rangle -\frac{i\chi }{2\zeta }\left( \sqrt{%
\frac{(n+2)!}{n!}}|n+2\rangle +\sqrt{\frac{n!}{(n-2)!}}|n-2\rangle \right) \\
&&-\frac{\chi ^{2}}{8\zeta ^{2}}\left( \sqrt{\frac{(n+4)!}{n!}}|n+4\rangle +%
\sqrt{\frac{n!}{(n-4)!}}|n-4\rangle \right)
\end{eqnarray*}%
The transition rate becomes%
\begin{eqnarray}
R_{n,n+2} &\approx &i\varepsilon \frac{\chi }{4\zeta }\sqrt{(n+1)(n+2)}\left[
(n+2)^{k}-n^{k}\right]  \notag \\
&=&i\varepsilon \frac{\chi }{\zeta }\sqrt{(n+1)(n+2)}M_{k}(n)  \label{m1}
\end{eqnarray}%
with $M_{2}(n)=n+1$ and $M_{3}(n)=3n^{2}/2+3n+2$. The eigenenergies read%
\begin{equation}
\lambda _{n}\approx \zeta n-\frac{\chi ^{2}}{\zeta }\left( 1+\frac{\chi ^{2}%
}{\zeta ^{2}}\right) (2n+1).  \label{m2}
\end{equation}%
So for the modulation frequency
\begin{equation}
\eta =|\lambda _{n+2}-\lambda _{n}|\approx 2|\zeta |\left[ 1-2\left( \chi
/\zeta \right) ^{2}-2\left( \chi /\zeta \right) ^{4}\right]  \label{giv}
\end{equation}%
all the eigenstates become resonantly coupled (apart from the small shifts
introduced by the neglect of rapidly oscillating terms \cite{AVD14}). In our
case $\zeta $ can be both positive and negative, therefore DCE takes place
whenever $\omega _{1}\pm \eta \approx 2\omega _{0}$. We verified that for $%
\chi n\ll |\zeta |$ the approximate expressions (\ref{m1}) -- (\ref{m2}) are
in excellent agreement with the exact numeric diagonalization of the
Hamiltonian (\ref{Hintfin}).

\begin{figure}[h]
\begin{center}
\includegraphics[width=0.49\textwidth]{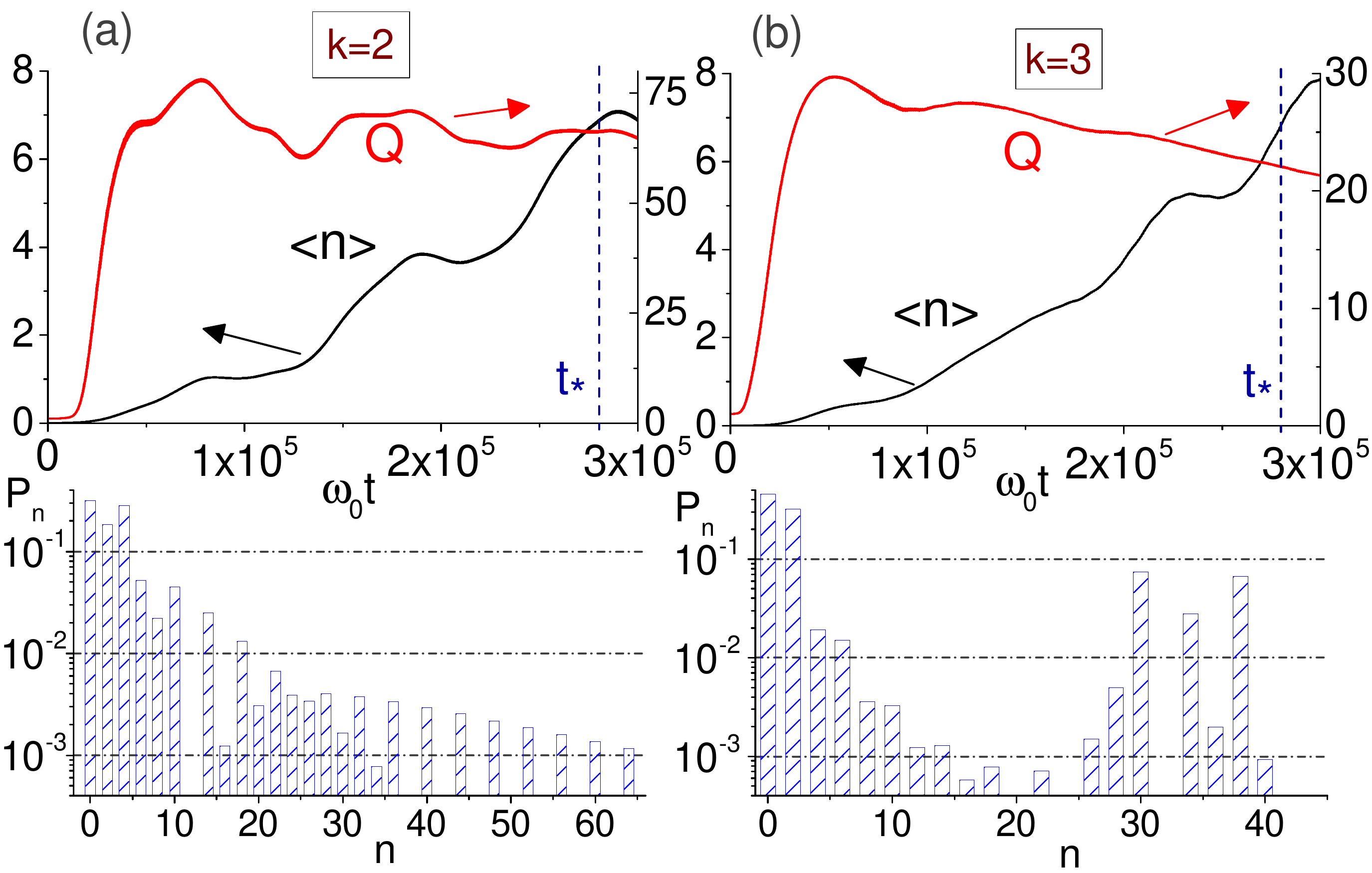} {}
\end{center}
\caption{(Color online) a) Simultaneous modulation of the Kerr nonlinearity
and the cavity frequency for parameters $\protect\omega _{1}=5\protect\omega %
_{0}$ and $\protect\eta \approx 2|\protect\omega _{0}-\protect\omega _{1}/2|$%
. Top panel: $\langle n\rangle $ and $Q$ as function of time. Bottom panel:
photon statistics at the time instant $t_{\ast }=2.8\times 10^{5}\protect%
\omega _{0}^{-1}$ (indicated by the vertical dashed line). b) Similar
analysis for the modulation of the third order nonlinearity and $\protect%
\omega _{1}=0.7\protect\omega _{0}$.}
\label{fig5}
\end{figure}

We solved numerically the Schr\"odinger equation corresponding to the
Hamiltonian (\ref{taine}). In Fig. \ref{fig5}a we illustrate the photon
generation from the initial vacuum state $|0\rangle $ for the modulation of
Kerr nonlinearity ($k=2$) with parameters $\omega _{1}=5\omega _{0}$, $%
\varepsilon _{\omega }=10^{-2}\omega _{0}$, $\varepsilon =10^{-3}\omega _{0}$
and $\eta $ given by Eq. (\ref{giv}). The panel on the top illustrates the
behavior of $\langle n\rangle $ and $Q$, while the panel on the bottom
illustrates the photon statistics at the time instant $\omega _{0}t_{\ast
}=2.8\times 10^{5}$. The long tail of the photon number distribution
explains the high value of the $Q$-factor. In Fig. \ref{fig5}b we repeat the
analysis for the modulation of the third-order nonlinearity with the
modified parameters $\omega _{1}=0.7\omega _{0}$ and $\eta =2|\zeta
|[1+4(\chi /\zeta )^{2}]$. This figure attests that photons can be generated
from vacuum via modulation of cavity nonlinearities even in the absence of
additional subsystems, provided the cavity frequency is also modulated.
In a sense, this is an interesting example of a positive ``interference''
between two different processes. Indeed, there is no photon generation in the standard
DCE configuration with a high detuning of the wall vibration frequency in the
absence of the Kerr medium. And no photon generation in the cavity at rest
in the presence of a single Kerr nonlinearity. However, when two mechanisms
are combined in a thoroughly thought way -- the generation becomes possible!
Another example of such kind of ``interference'' connected with the Kerr
medium was demonstrated in Ref. \cite{Chum95}.

\section{Discussion}
\label{sec-concl}

Our results can be easily extended to other subsystems coupled to the cavity
field and nonperiodic modulations, and they indicate that the photon
generation from vacuum due to modulation of the nonlinear terms is possible.
The peculiarity of this scheme is that the photon statistics is completely
different from the squeezed vacuum state, since the transition rates grow
much faster with $n$ than for the parametric amplification process. The main
disadvantages are the same as for other proposals of DCE, namely, the
requirement of a large modulation amplitude and a finely-tuned resonant
modulation for sufficiently long period of time. Since previous works
predicted the photon generation from vacuum in cavity or circuit QED due to
the modulation of cavity frequency, atomic frequency or atom--field coupling
strength, this work complements them by proving that under specific
circumstances the modulation of cavity nonlinearities can also be employed to
achieve generation of new states of light.
The statistics of these states is {\em hyper-Poissonian\/} ($Q \gg 1$), similar to the
states considered in Ref. \cite{DD-hyp}.
Although the mean number of photons generated from vacuum in the schemes discussed above
is not very big, probably, it could be increased after more thorough investigations
of possible experimental realizations in the circuit QED arrangements.
In any case, the discovery of a new kind of the parametric DCE-like effects in nonlinear
systems seems a significant achievement in the area of dynamical Casimir physics.
Note that quite recently other mechanisms of nonlinear DCE were considered in Ref.
\cite{Trun-arx}.

\begin{acknowledgments}
Partial support from National Council for Scientific and Technological
Development -- CNPq (Brazil) is acknowledged.
\end{acknowledgments}

\end{document}